\documentstyle[vsolj01,graphicx,natbib]{article}

\RequirePackage[T1]{fontenc}

\def\cite{\citealt}
\setcitestyle{aysep={}}

\def\Shibataprep{M. Shibata et al. in preparation}

\def\commenta{$^*$}
\def\commentb{$^\dagger$}
\def\commentc{$^\ddagger$}

\def\Pplus{P$_+\,$}

\begin{document}

\title{Periodic modulations during a long outburst in V363 Lyr}

\author{Taichi Kato$^1$}
\author{$^1$ Department of Astronomy, Kyoto University,
       Sakyo-ku, Kyoto 606-8502, Japan}
\email{tkato@kusastro.kyoto-u.ac.jp}

\begin{abstract}
I analyzed the Kepler long and short cadence data of V363 Lyr.
A period of 0.185723(8)~d was persistently detected and
this is identified as the orbital period.
V363 Lyr showed one long outburst accompanied by an
``(embedded) precursor'' during Kepler observations
and modulations with a period of 0.1956(2)~d, longer than
the orbital one, were detected during this outburst.
There are two possible interpretations of this period.
The first one is superhumps despite that V363 Lyr
is far above the period gap.  This interpretation requires
an evolved, undermassive secondary enabling a low mass ratio
of $q$=0.15.  The evolution of this long-period variations,
however, does not follow the standard evolution of superhumps.
The second one is that the precursor occurred when
the disk reached the tidal truncation radius, as inferred
from observations of IW And stars.  In this case,
the long-period variations could be interpreted as a variable
stream impact on a precessing eccentric disk, which
may have been formed by disturbances at the tidal truncation
radius.  This might lead to effective removal of
the angular momentum which resulted in 0.3--0.4 mag
brightening following the precursor.  The fractional
period excess suggests that $q$ is just above the stability
limit of the 3:1 resonance.  In either cases, 
the nature of the secondary and the mass ratio need to
be verified by spectroscopic observations.
\end{abstract}

\section{Introduction}

   V363 Lyr is a dwarf nova discovered by \citet{hof67an289205}.
\citet{gal85nearM56var} studied this object photographically.
Although both authors suggested high frequency of outbursts,
the interval ($\sim$21.4~d) was first confidently measured
by CCD observations by \citet{kat01v363lyr}. \citet{liu99CVspec1}
spectroscopically confirmed this object to be
a cataclysmic variable (CV).
This object is included in the Kepler field
\citep{bor10Keplerfirst,Kepler} and was
observed as a Kepler target as KIC 7431243.
\citet{ram14RATSKepler} reported an analysis of the 5.2~d
segment and detected two periods of 4.68~hr and 4.47~hr,
which I will discuss in this paper.

\section{Data Analysis}

   The data analysis was performed practically in the same way
as in \citet{osa13v1504cygKepler,osa13v344lyrv1504cyg}:
I used two-dimensional Fourier analysis for detecting
the signals and examined $O-C$ diagrams to detect variation
of the periods.

   Two sets of Kepler data for V363 Lyr are publicly available:
quarter 15 (Q15) short cadence (SC) and Q16 long cadence (LC) data.
Since the Q15 observations (in quiescence) covered only for 5.3~d, 
I primarily used this quarter for confirming the period detected 
in Q16.  The Q16 data contained five short outbursts and
one long outburst.

\section{Results}

\subsection{Identification of two periods}

\begin{figure*}
  \begin{center}
    \includegraphics[width=16cm]{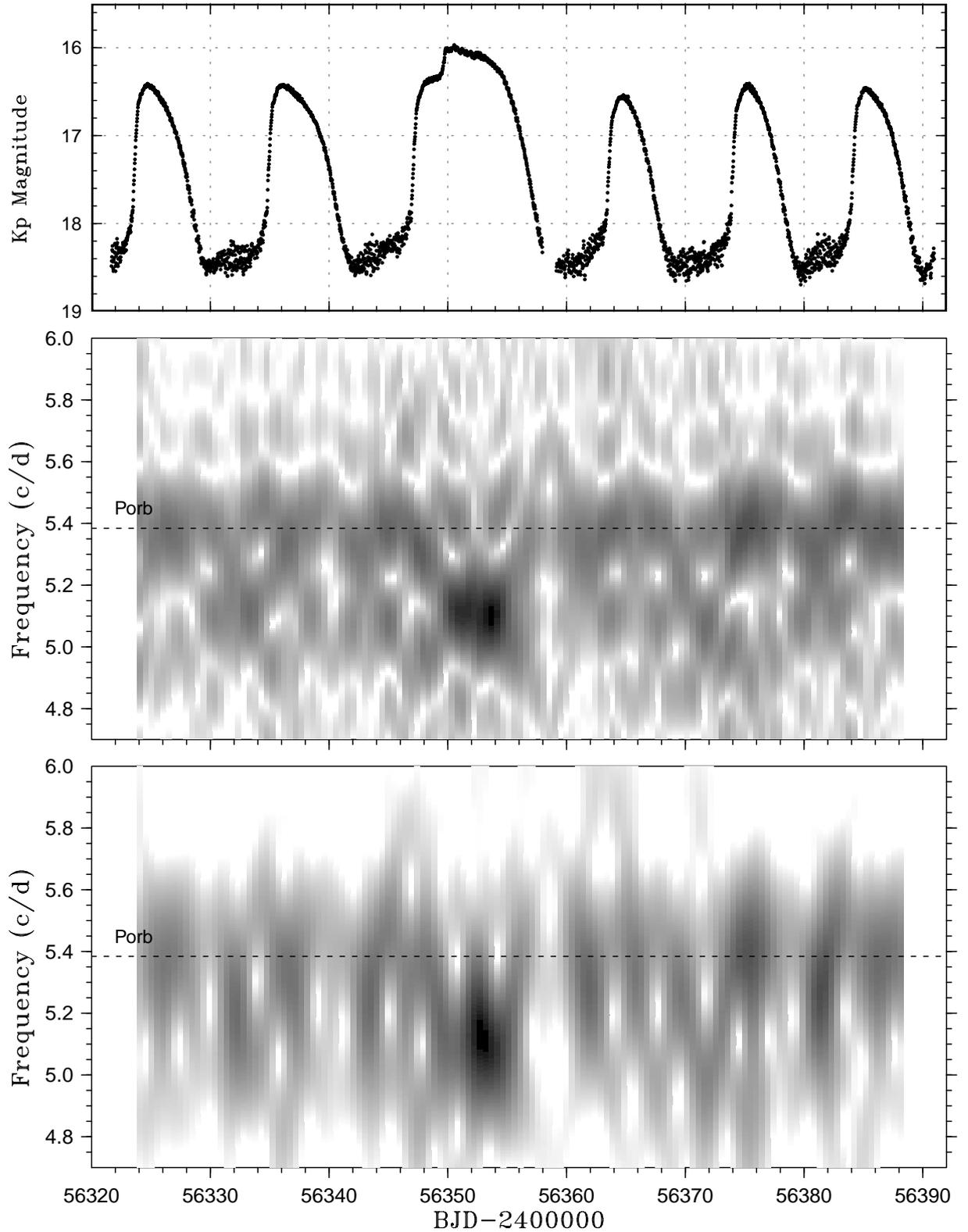}
  \end{center}
  \caption{Two-dimensional power spectrum of the Kepler long cadence
  light curve of V363 Lyr.
  (Upper:) Kepler Light curve.
  (Middle:) Power spectrum.  The width of 
  the sliding window and the time step used are 5~d and 0.5~d,
  respectively.  A Hanning window function was used.
  A signal (\Pplus) with a period longer than $P_{\rm orb}$
  appeared during the long outburst.
  (Lower) Two-dimensional PDM analysis.
  The width of the sliding window and the time step used
  are 5~d and 0.5~d, respectively.  Dark colors represent signals
  (lower $\theta$ in the PDM statistics).
  The orbital signal was present except during the long outburst.
  The signal of \Pplus was strongest during
  the long outburst.}
  \label{fig:v363spec2d}
\end{figure*}

   I present the result of period analysis only around
the frequencies of these periods.
There is no candidate frequency outside this frequency region.
The results of the Fourier analysis is shown
in figure \ref{fig:v363spec2d}. 
Two major signals were present.  These signals corresponds
to the periods detected by \citet{ram14RATSKepler}.

\begin{figure*}
  \begin{center}
    \includegraphics[width=16cm]{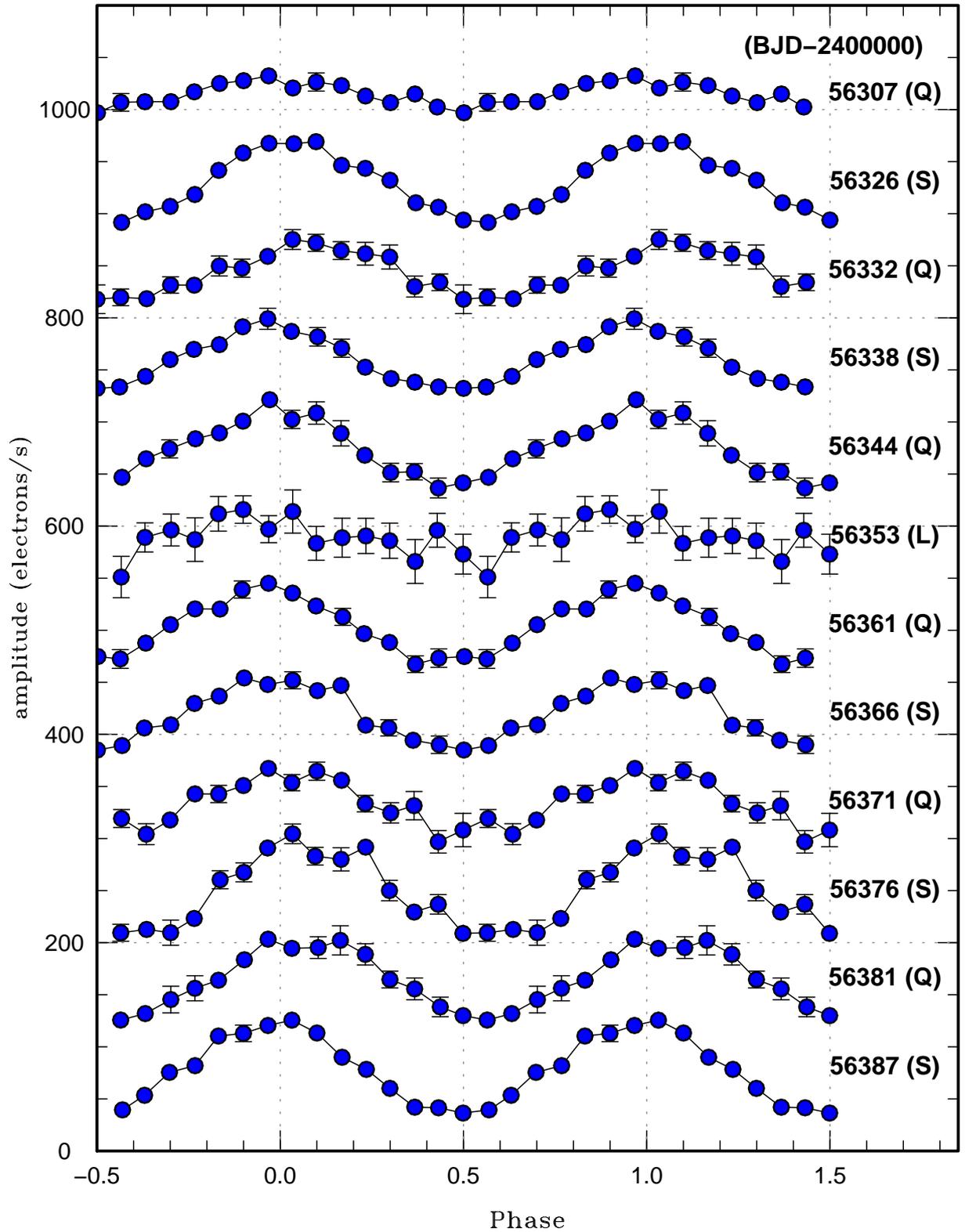}
  \end{center}
  \caption{Variation of the profile of the 5.38-c/d signal.
  Each phase-averaged light curve was constructed from
  4-d Kepler observations centered on the dates shown
  in the figure.  The phase$=0$ and period were defined
  as BJD 2456351.979 and 0.185723~d, respectively.
  The days 56307, 56332, 56344, 56361, 56371
  and 56381 correspond to quiescence (Q in the figure);
  the days 56326, 56338, 56366, 56376 and 56387 correspond
  to short outbursts (S).
  The day 56353 corresponds to the long outburst (L).
  Only except during the long outburst, the 5.38-c/d signal
  was detected with a constant phase and almost constant
  amplitudes (in flux).
  }
  \label{fig:v363prof}
\end{figure*}

\begin{figure*}
  \begin{center}
    \includegraphics[width=16cm]{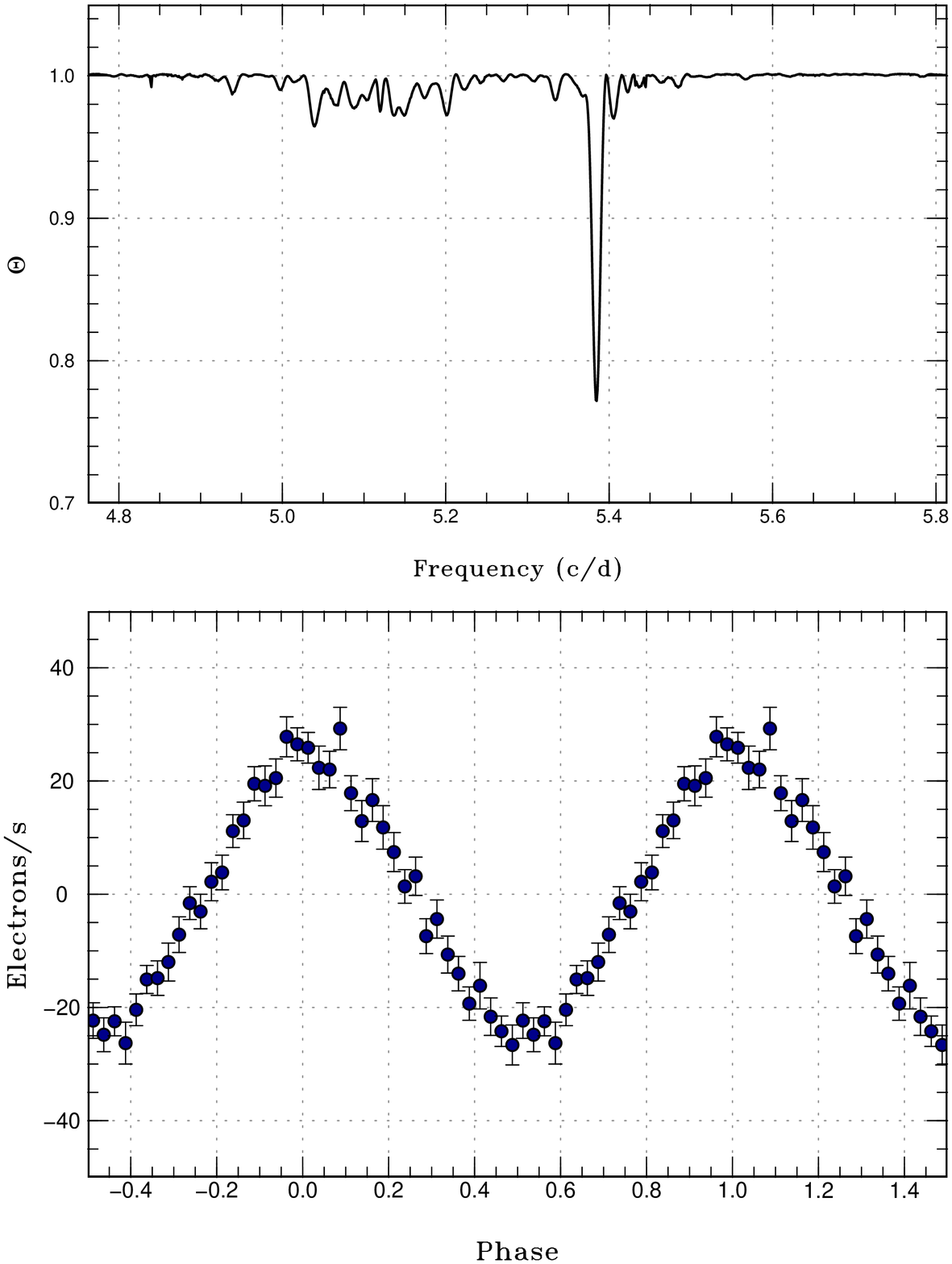}
  \end{center}
  \caption{PDM analysis of V363 Lyr of the entire data.
  Upper: PDM analysis.  The sharp signal at 0.18572~d
  is $P_{\rm orb}$.
  Lower: mean profile of the orbital signal.
  }
  \label{fig:v363porbpdm}
\end{figure*}

   I examined the stability of the signal at 5.38~c/d.
Almost all the segments (4~d in width), both in quiescence
and in outburst, showed the constant phase and almost
constant amplitude (in flux) of this 5.38-c/d signal
(figure \ref{fig:v363prof}).  I conclude that this signal
was coherent during the Kepler observation, and identified it 
to be the orbital period ($P_{\rm orb}$).  A phase dispersion
minimization (PDM: \cite{PDM}) analysis of the combined
data set of Q15 and Q16 yielded a period of 0.185723(8)~d
(figure \ref{fig:v363porbpdm}).  The error was estimated
by methods of \citet{fer89error} and \citet{Pdot2}.

   A two-dimensional Fourier analysis (figure \ref{fig:v363spec2d})
clearly indicates that another signal around frequencies 5.0--5.2~c/d
was especially strong around the long outburst.  Since the
frequency of this signal is lower (i.e. the period is longer)
than that of $P_{\rm orb}$.  Hereafter I call this signal \Pplus.
A two-dimensional PDM analysis is also shown in
figure \ref{fig:v363spec2d}.
In the case of V363 Lyr, PDM analysis gave a better
result than least absolute shrinkage and selection operator
(Lasso: \cite{Lasso,kat12perlasso,kat13j1924})
analysis employed in \citet{osa13v344lyrv1504cyg}
due to the non-sinusoidal nature of the signal
and small number of points in the LC data.  The resolution of
PDM analysis is intermediate between Fourier transform
and Lasso analysis, and this result can be treated as
a slightly degraded version of Lasso analysis presented
in \citet{osa13v344lyrv1504cyg} [see also an application
of a two-dimensional PDM analysis to superhumps
in \citet{kat21bocet}].  It looks like that \Pplus
was sometimes weakly excited outside the long outburst.
No evidence of the negative superhump was detected.

\subsection{Development of \Pplus}

   The long outburst had a shoulder (precursor)
followed by a brighter outburst.
This phenomenon in SS Cyg stars is also referred to as
an ``embedded precursor'' by \citet{can12ugemLC}.
During the precursor, humps recurring with $P_{\rm orb}$
were recorded.  After BJD 2456352 (two days after
the peak), \Pplus humps appeared.

\begin{figure*}
  \begin{center}
    \includegraphics[width=16cm]{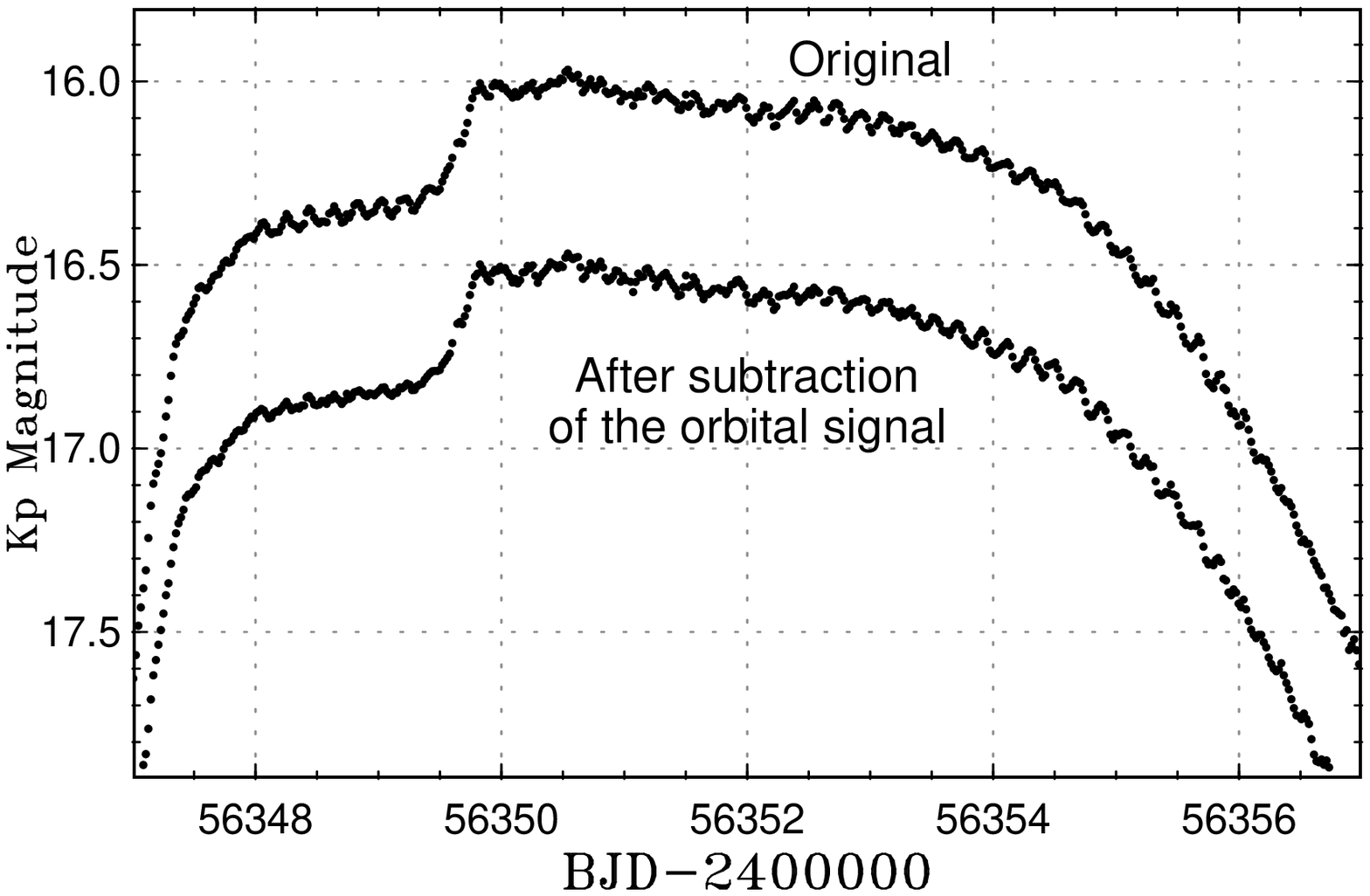}
  \end{center}
  \caption{Long outburst of V363 Lyr.
  The light curve of original observations (upper)
  and the light curve after subtraction (in flux) of
  the mean orbital signal are shown.
  The latter light curve was shifted by 0.5 mag.
  This long outburst had a shoulder (precursor)
  followed by a brighter outburst.
  Semi-regular variations were visible during the precursor,
  which can be interpreted as the orbital signal.
  Two days after the object reached the peak brightness,
  \Pplus humps appeared.
  }
  \label{fig:v363so}
\end{figure*}

\begin{figure*}
  \begin{center}
    \includegraphics[width=16cm]{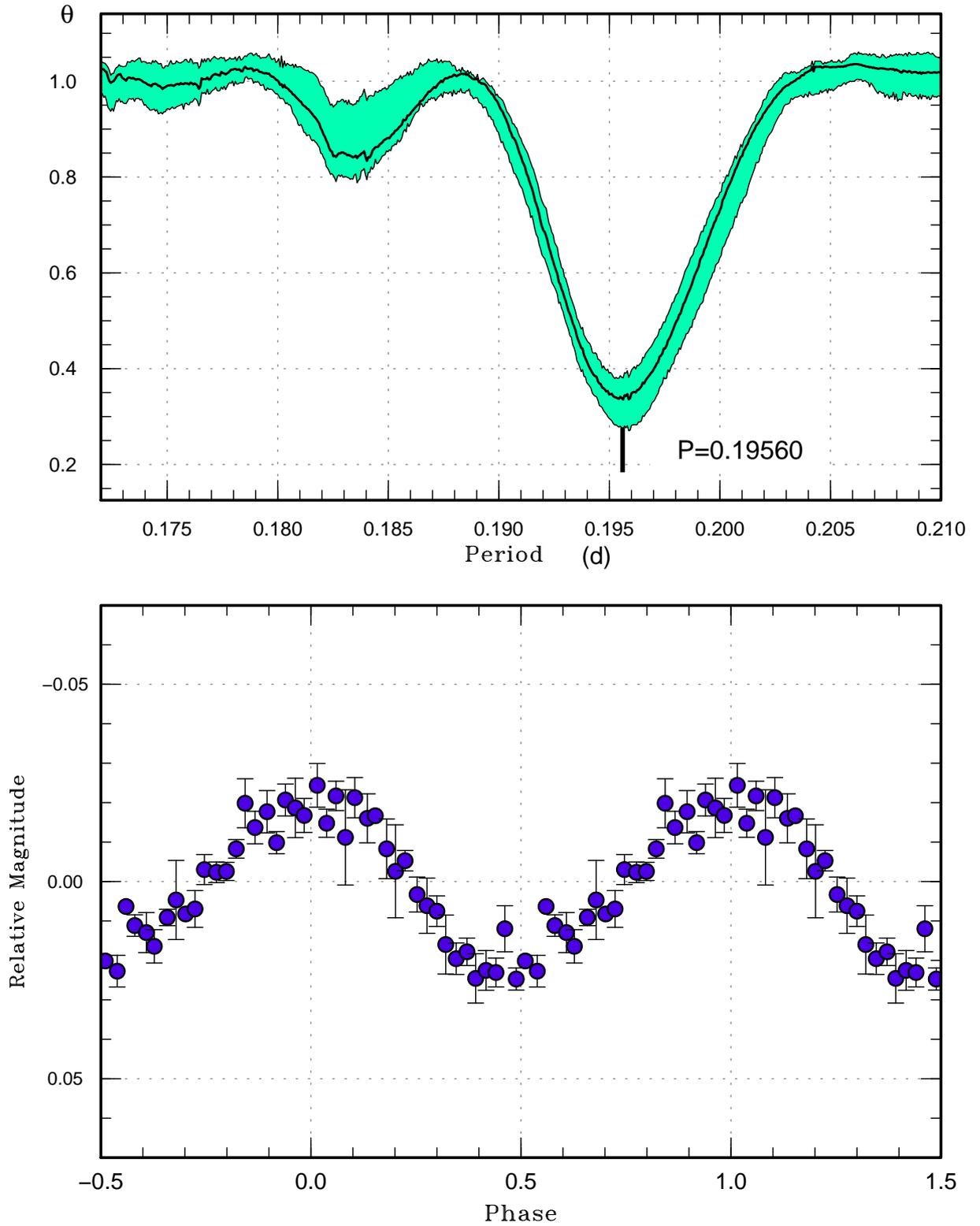}
  \end{center}
  \caption{PDM analysis of V363 Lyr during the long outburst
  when long-period variations were apparent
  (BJD 2456352--2456356.5).
  Upper: PDM analysis.
  I analyzed 100 samples which randomly contain 50\% of
  observations and the result is shown as a form of 90\%
  confidence intervals in the resultant PDM $\theta$ statistics.
  The signal at 0.1956~d corresponds to \Pplus.
  Lower: mean profile of \Pplus variations.
  }
  \label{fig:v363shpdm}
\end{figure*}

   I measured the times of maxima of these humps
in the light curve by the template fitting method described
in \citet{Pdot} after removing the trend of the outburst
by locally-weighted polynomial regression 
(LOWESS: \cite{LOWESS}).  I used the raw light curve
(without subtraction of the orbital signal) in
figure \ref{fig:v363so}.
The times of maxima are listed in table \ref{tab:shmax}.
As shown in figure \ref{fig:v363so}, the variations during
the precursor can be identified as the orbital signal.
After the precursor, the object brightened by
0.3--0.4 mag.  No clearly periodic variations were
detected for 2~d.  The amplitudes of \Pplus
were small ($\sim$0.04~mag) and there was
no significant variation of the period.
The profile of \Pplus was slightly
asymmetric (figure \ref{fig:v363shpdm}).

\begin{figure*}
  \begin{center}
    \includegraphics[width=16cm]{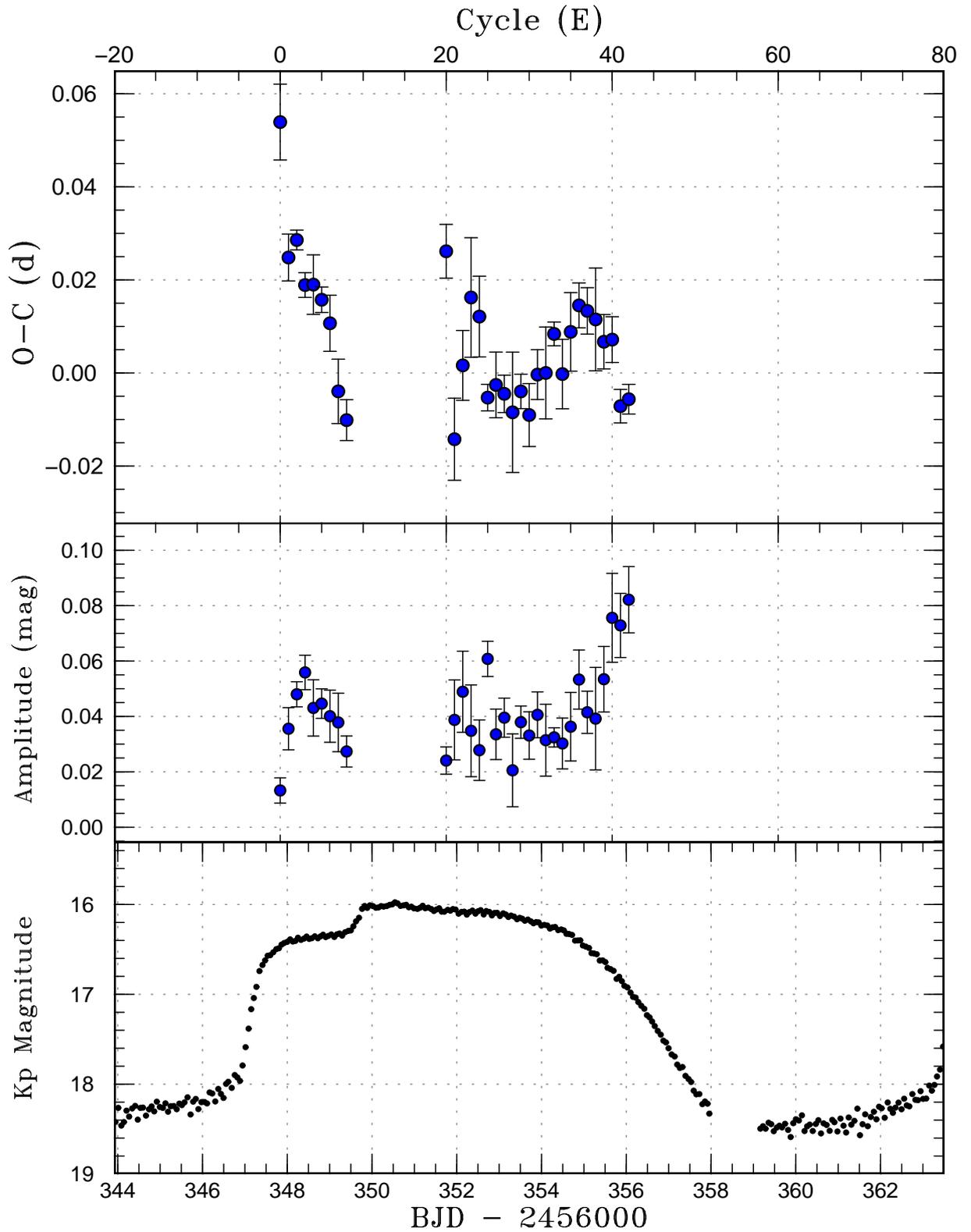}
  \end{center}
  \caption{$O-C$ diagram of humps during the long outburst
     in V363 Lyr.
     (Upper:) $O-C$ diagram.  The maxima for $E \le 8$
     refer to the orbital signal (see text).
     \Pplus appeared around $E=20$.
     I used a period of 0.19585~d for calculating the $O-C$ residuals.
     (Middle:) Amplitudes of \Pplus humps.
     (Lower:) Light curve.  The data were binned to 0.065~d.
  }
  \label{fig:v363humpoc}
\end{figure*}

\begin{table*}
\caption{Times of maxima in V363 Lyr during the long outburst}\label{tab:shmax}
\begin{center}
\begin{tabular}{rp{55pt}p{40pt}r@{.}lr}
\hline
\multicolumn{1}{c}{$E$} & \multicolumn{1}{c}{max\commenta} & \multicolumn{1}{c}{error} & \multicolumn{2}{c}{$O-C$\commentb} & \multicolumn{1}{c}{$N$\commentc} \\
\hline
0 & 56347.8869 & 0.0081 & 0&0351 & 8 \\
1 & 56348.0537 & 0.0050 & 0&0065 & 7 \\
2 & 56348.2533 & 0.0021 & 0&0107 & 7 \\
3 & 56348.4395 & 0.0027 & 0&0016 & 8 \\
4 & 56348.6354 & 0.0064 & 0&0022 & 8 \\
5 & 56348.8280 & 0.0027 & $-$0&0006 & 7 \\
6 & 56349.0188 & 0.0060 & $-$0&0051 & 8 \\
7 & 56349.2000 & 0.0069 & $-$0&0193 & 8 \\
8 & 56349.3897 & 0.0044 & $-$0&0249 & 7 \\
20 & 56351.7762 & 0.0058 & 0&0175 & 7 \\
21 & 56351.9316 & 0.0088 & $-$0&0224 & 8 \\
22 & 56352.1433 & 0.0075 & $-$0&0060 & 8 \\
23 & 56352.3538 & 0.0128 & 0&0091 & 8 \\
24 & 56352.5455 & 0.0087 & 0&0055 & 6 \\
25 & 56352.7240 & 0.0029 & $-$0&0114 & 8 \\
26 & 56352.9225 & 0.0071 & $-$0&0082 & 8 \\
27 & 56353.1165 & 0.0040 & $-$0&0096 & 7 \\
28 & 56353.3083 & 0.0129 & $-$0&0131 & 7 \\
29 & 56353.5087 & 0.0037 & $-$0&0081 & 8 \\
30 & 56353.6995 & 0.0068 & $-$0&0126 & 8 \\
31 & 56353.9040 & 0.0054 & $-$0&0034 & 8 \\
32 & 56354.1002 & 0.0099 & $-$0&0026 & 8 \\
33 & 56354.3044 & 0.0026 & 0&0063 & 8 \\
34 & 56354.4917 & 0.0075 & $-$0&0018 & 8 \\
35 & 56354.6966 & 0.0084 & 0&0078 & 8 \\
36 & 56354.8981 & 0.0048 & 0&0140 & 8 \\
37 & 56355.0928 & 0.0050 & 0&0133 & 8 \\
38 & 56355.2868 & 0.0110 & 0&0120 & 7 \\
39 & 56355.4779 & 0.0058 & 0&0077 & 8 \\
40 & 56355.6742 & 0.0049 & 0&0087 & 7 \\
41 & 56355.8557 & 0.0036 & $-$0&0052 & 7 \\
42 & 56356.0531 & 0.0032 & $-$0&0031 & 7 \\
\hline
  \multicolumn{6}{l}{\commenta BJD$-$2400000.} \\
  \multicolumn{6}{l}{\commentb Against max $= 2456347.8519 + 0.195341 E$.} \\
  \multicolumn{6}{l}{\commentc Number of points used to determine the maximum.} \\
\end{tabular}
\end{center}
\end{table*}

\section{Discussion}

\subsection{Comparison with superhumps in SU UMa stars}

   As we have seen, V363 Lyr showed both the orbital
signal and the \Pplus signal longer than $P_{\rm orb}$ during
the long, bright outburst.
These \Pplus variations may look similar to
superhumps in SU UMa stars, which have periods a few
percent longer than $P_{\rm orb}$ and are considered to
arise from the precession of an eccentric accretion disk
whose deformation is excited by the 3:1 resonance
(\cite{whi88tidal}; \cite{hir90SHexcess}; \cite{lub91SHa}).

   There are, however, a number of features different from
superhumps in ordinary SU UMa stars.  They are:

\begin{itemize}

\item The orbital period is too long [0.185723(8)~d]
for a typical SU UMa star
[see e.g. \citet{war95book} for classical statistics],
although some long-$P_{\rm orb}$ dwarf novae are
classified as SU UMa stars (see e.g. \cite{kat21asassn19ax}).

\item The ratio between the durations of the long outburst
(12~d for the entire outburst, 6~d after the long period appeared)
and other outbursts (5~d) is small compared to
other SU UMa stars.

\item Superhumps in SU UMa stars usually show $O-C$ variations
\citep{Pdot}.  \Pplus in V363 Lyr, however, did not show
such a variation.

\item In many SU UMa stars, superhumps start to appear
during the precursor \citep{osa13v1504cygKepler,osa13v344lyrv1504cyg}.
In V363 Lyr, \Pplus appeared much later
(2~d after the end of the precursor).

\end{itemize}

\subsection{Can \Pplus be growing superhumps?}\label{sec:stagea}

   In many SU UMa-type dwarf novae, superhumps start to
appear during the precursor (if there is a precursor) or
a few to several days after the start of
the superoutburst (if there is no precursor).
This reflects when the 3:1 resonance starts
to develop \citep{osa05DImodel}.
During this growing phase of superhumps, superhump
periods ($P_{\rm SH}$) are longer and this phase is referred to
as stage A \citep{Pdot}.  It is interpreted that
during the growing phase of superhumps, the eccentric
part of the disk is confined to the region near
the 3:1 resonance \citep{osa13v344lyrv1504cyg,kat13qfromstageA}
and that the period reflects
the dynamical precession rate at the 3:1 resonance.
As the region of the eccentric part reaches
the inner part of the disk, the pressure effect
slows down the precession rate (stage B).
In many SU UMa-type dwarf novae, stage B superhumps
are seen near the peaks of superoutbursts.

   \Pplus in V363 Lyr has a period of 0.1956(2)~d
(figure \ref{fig:v363shpdm}), which is 5.3\% longer
than $P_{\rm orb}$.  This period corresponds to
a fractional superhump excess in frequency
$\epsilon^* \equiv 1-P_{\rm orb}/P_{\rm SH}$ =
0.0514(10).  If we consider that \Pplus variations
in V363 Lyr reflect the dynamical precession rate
at the 3:1 resonance, this value corresponds to a mass ratio of
$q$=0.152(3) \citep{kat13qfromstageA}.
The mass of the secondary inferred from
this mass ratio is too small for an object with
$P_{\rm orb}=$0.18572~d.  The mass, $M_V$ and $M_K$
ofthe secondary for this period on the standard evolutionary
sequence of CVs are 0.39$M_\odot$, $+$10.2 and $+$5.9,
respectively \citep{kni06CVsecondary}.
Even a Chandrasekhar-mass white dwarf gives $q$=0.28,
which cannot explain the observation.

   There remains a possibility that the secondary
in V363 Lyr is undermassive for $P_{\rm orb}$.
Such a system should have an evolved secondary
such as QZ Ser \citep{tho02qzser},
CRTS J134052.1$+$151341 \citep{tho13j1340}
and ASASSN-18aan \citep{wak21asassn18aan}.
The quiescent absolute magnitudes of V363 Lyr are
$M_V$=$+$7.5 and $M_K$=$+$4.4 [using 2MASS \citep{2MASS}
and Gaia parallax \citep{GaiaEDR3}].
These values are brighter than those of
main-sequence stars on the standard evolutionary
sequence of CVs, and the secondary in V363 Lyr
may indeed be evolved.
For comparison, quiescent absolute magnitudes are
$M_V$=$+$10.5 and $M_K$=$+$7.2 for QZ Ser,
$M_V$=$+$8.4 and $M_K$=$+$5.9 for CRTS J134052.1$+$151341
and $M_V$=$+$8.4 and $M_K$=$+$6.0 for ASASSN-18aan.
If the secondary is indeed evolved in V363 Lyr,
\Pplus might be attributed to superhumps arising from
the 3:1 resonance.
If this is the case, the difference in the behavior
of superhumps from other SU UMa stars would require
an additional explanation since the corresponding $q$
value is normal for an SU UMa star and textbook evolution
of superhumps is expected.
Spectroscopic determination of the secondary type and
radial-velocity measurements are needed.

\subsection{Could \Pplus be stream impact on a precessing eccentric disk?}

   Since \Pplus was seen during the middle-to-late phase of
the outburst, this variation might be attributed to
(traditional) late superhumps
\citep{hae79lateSH,vog83lateSH,vanderwoe88lateSH},
which arise from the stream impact on an eccentric disk.

The dynamical precession rate of an eccentric disk can be
obtained by the method in \citet{hir90SHexcess,kat13qfromstageA}
and the disk radius can be estimated by equating
the theoretical precession rate with the observed $\epsilon^*$.
The results for various $q$ values are listed in
table \ref{tab:diskradius}.  I used the formula of
the tidal truncation radius
\begin{equation}
r_{\rm tidal} = \frac{0.60}{1+q}
\end{equation}
in \citet{pac77ADmodel}.
and Lubow-Shu (or circularization) radius
\begin{equation}
r_{\rm LS} = 0.488 q^{-0.464}
\end{equation}
from \citet{lub75AD}.
The value $q$=0.15 corresponds to subsection \ref{sec:stagea}
(assuming that \Pplus is growing or stage A superhumps).
The values $q$=0.25 or $q$=0.30 correspond to the limit
of the 3:1 tidal instability.  The value $q$=0.476 corresponds
to a secondary star on the standard evolutionary sequence
of CVs \citep{kni06CVsecondary} and an average-mass
white dwarf 0.82$M_{\odot}$ in CVs \citep{zor11SDSSCVWDmass}.

   In cases for $q < 0.30$ the radius of the 3:1 resonance
is inside $r_{\rm tidal}$ and the tidal instability due
to the 3:1 resonance can occur.  In these cases,
an eccentric disk is expected to form by the tidal instability
and the identification \Pplus as (traditional) late superhumps
is possible.  This interpretation, however,
requires an explanation why ordinary superhumps before
the appearance of late superhumps were not observed.
In all cases, the estimated radii are far outside $r_{\rm LS}$
and they are achievable.  For $q > 0.30$, however, eccentric
deformation of the disk by the 3:1 resonance is not expected
and there is a need for a different mechanism.

\begin{table*}
\caption{Disk radius estimated from the precession rate}\label{tab:diskradius}
\begin{center}
\begin{tabular}{ccccc}
\hline
$q$   & Disk radius\commenta & Truncation radius\commenta & Radius of 3:1 resonance\commenta & Lubow-Shu radius\commenta \\
\hline
0.15  & 0.462 & 0.52 & 0.459 & 0.118 \\
0.20  & 0.413 & 0.50 & 0.453 & 0.103 \\
0.25  & 0.377 & 0.48 & 0.447 & 0.093 \\
0.30  & 0.348 & 0.46 & 0.441 & 0.085 \\
0.35  & 0.325 & 0.44 & 0.435 & 0.079 \\
0.476\commentb & 0.283 & 0.41 & 0.422 & 0.069 \\
\hline
  \multicolumn{4}{l}{\commenta Unit: binary separation.} \\
  \multicolumn{4}{l}{\commentb Assuming a standard secondary and an average mass white dwarf.} \\
\end{tabular}
\end{center}
\end{table*}

\subsection{Possible deformation of the disk for a high-$q$ system}

   If the true mass ratio is too large to hold
the 3:1 resonance within the tidal truncation radius,
V363 Lyr is not an SU UMa star.  \citet{kat21stcha} recently
proposed an interpretation that embedded precursors in
long outbursts of SS Cyg stars may correspond to a phenomenon when
the disk radius reaches the tidal truncation radius.
This interpretation originated from direct observations
of the variation of the disk radius (\Shibataprep)
in an IW And star \citep{sim11zcamcamp1,kat19iwandtype}.
If this is indeed the case in the precursor
in V363 Lyr, \Pplus may be a previously undescribed phenomenon
which is excited when the disk reaches the tidal truncation radius.
In IW And stars, standstills are terminated by
brightening and this would require a mechanism of
effective removal of angular momentum
at the tidal truncation radius
(see \cite{kim20kic9406652}; \cite{kat21bocet};
\Shibataprep).  It would not be surprising if
intersections of orbits in the disk near
the tidal truncation radius
\citep{pac77ADmodel} cause disturbances and lead to
eccentric deformation of the disk (which is desired
for the late-superhump type interpretation) and
effective removal of the angular momentum.
As seen in table \ref{tab:diskradius}, the estimated
disk radius is much smaller than $r_{\rm tidal}$ for
larger $q$ values.  This mechanism to produce \Pplus
would only be acceptable in V363 Lyr
for a narrow range of $q$ which does not allow
the 3:1 resonance to occur but enables the disk radius
sufficiently close to $r_{\rm tidal}$.
Brightening by 0.3--0.4 mag after the precursor
in V363 Lyr is probably the result of effective removal
of the angular momentum (just as in superoutburst in
SU UMa stars, \cite{osa89suuma}), and
\Pplus observed in V363~Lyr could be a manifestation of
the disturbances around the tidal truncation radius.

\section*{Acknowledgments}

I thank the Kepler Mission team and the data calibration engineers for
making Kepler data available to the public.
This work was supported by JSPS KAKENHI Grant Number 21K03616.

\newcommand{\noop}[1]{}


\begin{thebibliography}{}

\bibitem[{Borucki} et~al.(2010)]{bor10Keplerfirst}
  {Borucki}, W.~J. {et~al.} (2010) {Kepler} planet-detection mission:
  Introduction and first results. {\em Science\,} {\bf 327}, 977

\bibitem[{Cannizzo}(2012)]{can12ugemLC}
  {Cannizzo}, J.~K. (2012) The shape of long outbursts in {U Gem} type dwarf
  novae from {AAVSO} data. {\em ApJ\,} {\bf 757}, 174

\bibitem[{Cleveland}(1979)]{LOWESS}
  {Cleveland}, W.~S. (1979) {Robust locally weighted regression and smoothing
  scatterplots}. {\em J. Amer. Statist. Assoc.\,} {\bf 74}, 829

\bibitem[{Cutri} et~al.(2003)]{2MASS}
  {Cutri}, R.~M. {et~al.} (2003) {2MASS} {All Sky Catalog} of point sources
  (NASA/IPAC Infrared Science Archive)

\bibitem[{Fernie}(1989)]{fer89error}
  {Fernie}, J.~D. (1989) Uncertainties in period determinations. {\em PASP\,}
  {\bf 101}, 225

\bibitem[{Gaia Collaboration} et~al.(2021)]{GaiaEDR3}
  {Gaia Collaboration} {et~al.} (2021) {Gaia Early Data Release} 3. summary of
  the contents and survey properties. {\em A\&A\,} {\bf 649}, A1

\bibitem[Galkina and Shugarov(1985)]{gal85nearM56var}
  Galkina, M.~P., \& Shugarov, S.~{\relax Yu.} (1985) Study of 11 stars in
  {Lyra}. {\em Perem.\ Zvezdy\,} {\bf 22}, 225

\bibitem[Haefner et~al.(1979)]{hae79lateSH}
  Haefner, R., Schoembs, R., \& Vogt, N. (1979) The outbursts of the dwarf nova
  {VW Hydri} -- a comparative study of short and long eruptions. {\em A\&A\,}
  {\bf 77}, 7

\bibitem[{Hirose} and {Osaki}(1990)]{hir90SHexcess}
  {Hirose}, M., \& {Osaki}, Y. (1990) Hydrodynamic simulations of accretion
  disks in cataclysmic variables -- superhump phenomenon in {SU UMa} stars.
  {\em PASJ\,} {\bf 42}, 135

\bibitem[{Hoffmeister}(1967)]{hof67an289205}
  {Hoffmeister}, C. (1967) Mitteilungen {\"{u}ber} neuentdeckte
  {ver\"{a}nderliche} {Sterne}. {\em Astron.\ Nachr.\,} {\bf 289}, 205

\bibitem[{Kato}(2019)]{kat19iwandtype}
  {Kato}, T. (2019) Three {Z Cam}-type dwarf novae exhibiting {IW And}-type
  phenomenon. {\em PASJ\,} {\bf 71}, 20

\bibitem[{Kato} and {Hambsch}(2021)]{kat21stcha}
  {Kato}, T., \& {Hambsch}, F.-J. (2021) On the nature of embedded precursors
  in long outbursts of {SS Cyg} stars as inferred from observations of the {IW
  And} star {ST Cha}. {\em VSOLJ\ Variable\ Star\ Bull.\,} {\bf 83},
  (arXiv:2110.10321)

\bibitem[{Kato} et~al.(2009)]{Pdot}
  {Kato}, T. {et~al.} (2009) Survey of period variations of superhumps in {SU
  UMa}-type dwarf novae. {\em PASJ\,} {\bf 61}, S395

\bibitem[{Kato} et~al.(2021a)]{kat21asassn19ax}
  {Kato}, T. {et~al.} (2021a) {ASASSN-19ax}: {SU UMa}-type dwarf nova with a
  long superhump period and post-superoutburst rebrightenings. {\em VSOLJ\
  Variable\ Star\ Bull.\,} {\bf 84}, (arXiv:2111.01304)

\bibitem[{Kato} and {Maehara}(2013)]{kat13j1924}
  {Kato}, T., \& {Maehara}, H. (2013) Analysis of {Kepler} light curve of the
  novalike cataclysmic variable {KIC 8751494}. {\em PASJ\,} {\bf 65}, 76

\bibitem[{Kato} et~al.(2010)]{Pdot2}
  {Kato}, T. {et~al.} (2010) {Survey of Period Variations of Superhumps in {SU
  UMa}-Type Dwarf Novae. {II}. The Second Year (2009-2010)}. {\em PASJ\,} {\bf
  62}, 1525

\bibitem[Kato et~al.(2001)]{kat01v363lyr}
  Kato, T., Nogami, D., Baba, H., \& Masuda, S. (2001) Outburst cycle of {V363
  Lyr}. {\em IBVS\,} {\bf 5118}

\bibitem[{Kato} and {Osaki}(2013)]{kat13qfromstageA}
  {Kato}, T., \& {Osaki}, Y. (2013) {\noop{KatoOsaki2013b}} {New} method to
  estimate binary mass ratios by using superhumps. {\em PASJ\,} {\bf 65}, 115

\bibitem[{Kato} et~al.(2021b)]{kat21bocet}
  {Kato}, T. {et~al.} (2021b) {BO Ceti}: Dwarf nova showing both {IW And} and
  {SU UMa}-type features. {\em PASJ\,}  in press (arXiv:2106.15028)

\bibitem[{Kato} and {Uemura}(2012)]{kat12perlasso}
  {Kato}, T., \& {Uemura}, M. (2012) Period analysis using the {Least Absolute
  Shrinkage and Selection Operator} ({Lasso}). {\em PASJ\,} {\bf 64}, 122

\bibitem[{Kimura} et~al.(2020)]{kim20kic9406652}
  {Kimura}, M., {Osaki}, Y., \& {Kato}, T. (2020) {KIC 9406652}: A laboratory
  of the tilted disk in cataclysmic variable stars. {\em PASJ\,} {\bf 72}, 94

\bibitem[{Knigge}(2006)]{kni06CVsecondary}
  {Knigge}, C. (2006) The donor stars of cataclysmic variables. {\em MNRAS\,}
  {\bf 373}, 484

\bibitem[{Koch} et~al.(2010)]{Kepler}
  {Koch}, D.~G. {et~al.} (2010) {Kepler} mission design, realized photometric
  performance, and early science. {\em ApJ\,} {\bf 713}, L79

\bibitem[Liu et~al.(1999)]{liu99CVspec1}
  Liu, Wu., Hu, J.~Y., Zhu, X.~H., \& Li, Z.~Y. (1999) Spectroscopic
  confirmation of 55 northern and equatorial cataclysmic variables. {I}. 27
  confirmed cataclysmic variables. {\em ApJS\,} {\bf 122}, 243

\bibitem[{Lubow}(1991)]{lub91SHa}
  {Lubow}, S.~H. (1991) {\noop{Lubow1991a}} {A} model for tidally driven
  eccentric instabilities in fluid disks. {\em ApJ\,} {\bf 381}, 259

\bibitem[{Lubow} and {Shu}(1975)]{lub75AD}
  {Lubow}, S.~H., \& {Shu}, F.~H. (1975) Gas dynamics of semidetached binaries.
  {\em ApJ\,} {\bf 198}, 383

\bibitem[{Osaki}(1989)]{osa89suuma}
  {Osaki}, Y. (1989) A model for the superoutburst phenomenon of {SU Ursae
  Majoris} stars. {\em PASJ\,} {\bf 41}, 1005

\bibitem[{Osaki}(2005)]{osa05DImodel}
  {Osaki}, Y. (2005) The disk instability model for dwarf nova outbursts. {\em
  Proc.\ Japan\ Acad.\ Ser.\ B\,} {\bf 81}, 291

\bibitem[{Osaki} and {Kato}(2013a)]{osa13v1504cygKepler}
  {Osaki}, Y., \& {Kato}, T. (2013a) {\noop{OsakiKato2013a}} {The} cause of the
  superoutburst in {SU UMa} stars is finally revealed by {Kepler} light curve
  of {V1504 Cygni}. {\em PASJ\,} {\bf 65}, 50

\bibitem[{Osaki} and {Kato}(2013b)]{osa13v344lyrv1504cyg}
  {Osaki}, Y., \& {Kato}, T. (2013b) {\noop{OsakiKato2013b}} {Study} of
  superoutbursts and superhumps in {SU UMa} stars by the {Kepler} light curves
  of {V344 Lyrae} and {V1504 Cygni}. {\em PASJ\,} {\bf 65}, 95

\bibitem[Paczy\'{n}ski(1977)]{pac77ADmodel}
  Paczy\'{n}ski, B. (1977) A model of accretion disks in close binaries. {\em
  ApJ\,} {\bf 216}, 822

\bibitem[{Ramsay} et~al.(2014)]{ram14RATSKepler}
  {Ramsay}, G. {et~al.} (2014) {RATS-Kepler} -- a deep high-cadence survey of
  the {Kepler} field. {\em MNRAS\,} {\bf 437}, 132

\bibitem[{Simonsen}(2011)]{sim11zcamcamp1}
  {Simonsen}, M. (2011) The {Z CamPaign}: Year 1. {\em J.\ American\ Assoc.\
  Variable\ Star\ Obs.\,} {\bf 39}, 66

\bibitem[Stellingwerf(1978)]{PDM}
  Stellingwerf, R.~F. (1978) Period determination using phase dispersion
  minimization. {\em ApJ\,} {\bf 224}, 953

\bibitem[{Thorstensen}(2013)]{tho13j1340}
  {Thorstensen}, J.~R. (2013) {CSS J134052.0$+$151341}: A cataclysmic binary
  star with a stripped, evolved secondary. {\em PASP\,} {\bf 125}, 506

\bibitem[Thorstensen et~al.(2002)]{tho02qzser}
  Thorstensen, J.~R., Fenton, W.~H., Patterson, J.~O., Kemp, J., Halpern, J.,
  \& Baraffe, I. (2002) {QZ Serpentis}: A dwarf nova with a 2-hour orbital
  period and an anomalously hot, bright secondary star. {\em PASP\,} {\bf 114},
  1117

\bibitem[{Tibshirani}(1996)]{Lasso}
  {Tibshirani}, R. (1996) Regression shrinkage and selection via the lasso.
  {\em J. R. Statistical Soc. Ser. B\,} {\bf 58}, 267

\bibitem[{van der Woerd} et~al.(1988)]{vanderwoe88lateSH}
  {van der Woerd}, H., {van der Klis}, M., {van Paradijs}, J., {Beuermann}, K.,
  \& {Motch}, C. (1988) Observations of the late superhump in {VW Hydri}. {\em
  ApJ\,} {\bf 330}, 911

\bibitem[{Vogt}(1983)]{vog83lateSH}
  {Vogt}, N. (1983) {VW Hydri} revisited -- conclusions on dwarf nova outburst
  models. {\em A\&A\,} {\bf 118}, 95

\bibitem[{Wakamatsu} et~al.(2021)]{wak21asassn18aan}
  {Wakamatsu}, Y. {et~al.} (2021) {ASASSN-18aan}: An eclipsing {SU UMa}-type
  cataclysmic variable with a 3.6-hr orbital period and a late {G}-type
  secondary star. {\em PASJ\,} {\bf 73}, 1209

\bibitem[Warner(1995)]{war95book}
  Warner, B. (1995) Cataclysmic Variable Stars (Cambridge: Cambridge University
  Press)

\bibitem[Whitehurst(1988)]{whi88tidal}
  Whitehurst, R. (1988) Numerical simulations of accretion disks. {I} -
  superhumps - a tidal phenomenon of accretion disks. {\em MNRAS\,} {\bf 232},
  35

\bibitem[{Zorotovic} et~al.(2011)]{zor11SDSSCVWDmass}
  {Zorotovic}, M., {Schreiber}, M.~R., \& {G{\"a}nsicke}, B.~T. (2011) Post
  common envelope binaries from {SDSS}. {XI}. the white dwarf mass
  distributions of {CVs} and pre-{CVs}. {\em A\&A\,} {\bf 536}, A42

\end{thebibliography}
\end{document}